\pdfoutput=1
\documentclass{article}

\usepackage[utf8]{inputenc} % allow utf-8 input
\usepackage[T1]{fontenc}    % use 8-bit T1 fonts
\PassOptionsToPackage{hyphens}{url}\usepackage{hyperref}       % hyperlinks
\usepackage{url}            % simple URL typesetting
\usepackage{booktabs}       % professional-quality tables
\usepackage{amsfonts}       % blackboard math symbols
\usepackage{nicefrac}       % compact symbols for 1/2, etc.
\usepackage{microtype}      % microtypography
\usepackage{lipsum}
\usepackage{apacite}

\title{Could regulating the creators deliver trustworthy AI?}

\date{}

\author{Labhaoise N{\'i} Fhaol{\'a}in and Andrew Hines$^{1,2}$\\
$^{1}$ School of Computer Science, University College Dublin, Ireland\\
$^{2}$ Insight Centre for Data Analytics, Ireland\\
labhaoise.ni.fhaolain@gmail.com, andrew.hines@ucd.ie}

\begin{document}
\maketitle

\textit{Is a new regulated profession, such as Artificial Intelligence (AI) Architect who is responsible and accountable for AI outputs necessary to ensure trustworthy AI?}

AI is becoming all pervasive and is often deployed in everyday technologies, devices and services without our knowledge. There is heightened awareness of AI in recent years which has brought with it fear. This fear is compounded by the inability to point to a trustworthy source of AI, however even the term "trustworthy AI" itself is troublesome. Some consider trustworthy AI to be that which complies with relevant laws, while others point to the requirement to comply with ethics and standards (whether in addition to or in isolation of the law). This immediately raises questions of whose ethics and which standards should be applied and whether these are sufficient to produce trustworthy AI in any event.

Various codes are used in and applied to the industry, with varying levels of enforceability  --- from the simple "don't be evil" mantra in Google's code of conduct through to ethical guidelines, best practices (e.g. EPSRC), standards such as IEEE \cite{Chatila2019TheSystems}, BS 8611 \cite{BritishStandardsInstitution2016BSSystems} with incorporated EPSRC principles of robotics \cite{2010PrinciplesWebsite}, laws (CE mark -- EU product safety), treaties (international trade rules or bilateral treaties of product/health standards). The lack of enforceability of rules and codes is not in citizens' or consumers' interests, or in the industry's interests as they must ensure consumers' rights are respected and protected.   

Aside from ethics and standards, when we apply law to AI, the speed at which the technology is evolving and the inconsequence of jurisdictional boundaries to this type of technology means that the legislature and the judiciary cannot adequately protect rights. 

\subsubsection*{"Regulation” of AI?}
Various approaches have emerged in the attempt to regulate AI -- through ethics codes, standards and law -- whether applying existing law or drafting AI specific statutes. Many commentators call for differing levels of intervention, dependent on the impact of the particular AI solution on society. When it comes to AI specific statutes, \citeA{Petit2017LawImplications} debates whether AI tailored legislation should be adopted for AI and robots or whether existing laws should be “repurposed” and applied, effectively by the courts, after the damage has occurred. His proposed framework operates on a scale ranging from full reliance on existing laws through to drafting new legislation. According to the author, the position on the scale depends on broader societal impact of the AI. 

This approach gives rise to the question of whether AI can be subject to effective legislative input in any event. In the context of unpredictability and uncontrollability of AI, and the call for increased transparency, \citeA{Buiten2019TowardsIntelligence} discusses whether explanations, which programmers can provide for algorithms, would be useful in certain legal scenarios. She concludes that the level of transparency required by law should depend on the impact of the AI solution on the individual.

While sliding scales are often the most equitable manner to apply law, it is troublesome when dealing with AI solutions -- often the impact is not known in advance and, after the damage has occurred, there is no way of re-setting the clock either for an individual or for the wider public.

An alternative to allowing the courts or tribunals to address breaches of legislation after breaches have occurred is to use standards and ethics codes to ensure that technology is being applied in the good of society. \citeA{Winfield2019EthicalAI} draws a distinction between the new generation of ethical standards and the conventional standards. Examining ten sets of ethical principals in existence in 2017 for AI and Robotics the author notes the urgent need for ethical principals to be converted into ethical standards. He notes that ACM and IEEE have recently published codes of ethics and professional conduct and calls for soft governance, giving the example of contracts which require compliance with standards as a condition of awarding procurement contracts. 

Even where there is legislation in place to govern AI, it may still be necessary to ensure compliance through a bottom up approach. \citeA{Mitrou2019DataIntelligence-Proof} uses the example of the GDPR in this regard.  The author questions whether the current data protection framework under the GDPR is “AI proof”. Noting that Data Protection laws are rooted in ethical considerations drawn from fundamental privacy rights and that a responsible controller must adhere not only to the text of the GDPR but also to its spirit, the author suggests that new accountability tools are necessary to ensure compliance.

\subsubsection*{Why not rely on ethics codes and standards?}

How to bridge the gap between statute and compliance is an age-old criminal and civil law conundrum. \citeA{Garvin1983CanWork} considered examples of industry self-regulation through standards in the absence of legislative input. As a result of the study, the author concludes that the best practice to achieve the best societal outcome was to use a two pronged approach of industry standard setting alongside government oversight.  \citeA{Huising2011GoverningRegulation} identify the ubiquitous gap between the enacted statutes and the statutes in action and that industry standards attempt to bridge the gap between the two. By studying the attempts to use management to regulate conduct, the authors conclude that compliance is not assured by the management system per se but rather through the individuals’ action and acceptance of responsibility.  

The above suggests that given the nature of the technology of AI, regardless of the levels of regulation and standards, some form of widespread policing is required to bring about the desirable outcome -- trustworthy AI. Whether this gap can be bridged through the use of a regulated profession to allow a bottom up approach to compliance should be explored.

\subsubsection*{Professional Regulation}
The concept of system of professions, the overlapping of areas of expertise and the changes in the nature of professions and the manner in which professions are regulated have been examined by 
\citeA{Abbott1988TheLabor, Gorman2011GoldenWork, Burns2019BeyondProfessions}. The reasons for the emergence of the regulated professions remains true today. It is in the public’s interest to ensure that when services are provided in areas which can have significance impact on citizens and consumers, high standards should be maintained and this is achieved through a bottom up approach, albeit with imposed regulation. This regulates the individual's conduct rather than regulating the product or service.

Many individuals and bodies would like their occupation to be recognised as a “profession”, which is generally the first step to achieving a regulated title. As an example, \citeA{Vinarova2010TheInformatician} seek to demonstrate the significance of “medical informatician” as an expert. In order to do so, they present the necessity of such experts in the medical field and establish the minimum required to establish a profession, being professional culture, skills, competencies and types of information required. Formal regulation may be the ultimate goal for an occupation and \citeA{Lester2016TheCommunities} presents four examples of self-regulated groups which developed without the support of any statutory framework. The author notes that the state as an entity has greater or lesser interest in the formal regulation of professions depending on the public interest aspect of the activities. Increasingly, the state has stepped in to govern regulation where the self-regulation has failed and this has led to “regulated self-regulation”.

Given the potential for harm to individuals by AI solutions, there is an argument that there is a great public interest in controlling certain activities of computer scientists in the same way that the state controls the activities of healthcare providers, accountants and lawyers.

Some commentators have examined the challenges facing the introduction of a specified profession within the computer science industry including conflict both interprofessionally on how roles might be distributed and how the interests of stakeholders are addressed~\cite{Holmes2000FashioningProfession,Holmes2010TheProfession,Adams2007InterprofessionalCanada,Godfrey2009SystemsProfession}.

\subsubsection*{Conclusion}
The creation of a regulated profession -- AI architect -– could be a solution to the issue of lack of trustworthy AI. The regulated professions are those whose actions can have a significant impact on the individual and on society -- lawyers and doctors were amongst the first regulated professions -- as these professionals' actions can impact the rights and health of individuals. Arguably the Tech industry can have an even more detrimental impact on individuals' rights, health and much more, yet the "practitioners" are not regulated. Appropriate and effective action is required to ensure that people and society are protected. While the regulated professions system is not without fault, it is a system which requires personal responsibility. While the legislature moves slowly, a regulated profession's code can be amended quickly which means that changes in technology can be taken account of in a timely manner.  

By taking a bottom up approach and placing the responsibility on the individuals, the ethical/legal/trustworthy 'by design' concept gains a foothold. The onus is placed on the individual to ensure that the product or service being produced is of a required standard, to "certify" the product in a manner similar to a chartered civil engineer. Currently there is an imbalance between the individual software engineer employee and the company. The individual employee who wants to highlight unethical or illegal practices in the development of AI solutions has very few options. If however, an AI Architect were accredited by a regulatory body and a company required certification from their AI Architect to apply the "Trustworthy AI" stamp, then the individual is in a more secure position. The AI Architect must comply with their professional obligations to maintain their professional accreditation and failing to comply could open them up to disciplinary and court action.  

The concept of computer science as a profession is not new, though state regulation of the occupation does not appear to have been canvassed to a detailed level. In particular the suggestion to regulate the role of AI architect has not been treated. This warrants further research due to the near universal desire for trustworthy AI. 

\bibliographystyle{apacite}

\begin{thebibliography}{}

\bibitem [\protect \citeauthoryear {%
Abbott%
}{%
Abbott%
}{%
{\protect \APACyear {1988}}%
}]{%
Abbott1988TheLabor}
\APACinsertmetastar {%
Abbott1988TheLabor}%
\begin{APACrefauthors}%
Abbott, A\BPBI D.%
\end{APACrefauthors}%
\unskip\
\newblock
\APACrefYear{1988}.
\newblock
\APACrefbtitle {{The system of professions : an essay on the division of expert
  labor}} {{The system of professions : an essay on the division of expert
  labor}}\ (\PrintOrdinal{1st Edition}\ \BEd).
\newblock
\APACaddressPublisher{}{University of Chicago Press}.
\PrintBackRefs{\CurrentBib}

\bibitem [\protect \citeauthoryear {%
Adams%
}{%
Adams%
}{%
{\protect \APACyear {2007}}%
}]{%
Adams2007InterprofessionalCanada}
\APACinsertmetastar {%
Adams2007InterprofessionalCanada}%
\begin{APACrefauthors}%
Adams, T\BPBI L.%
\end{APACrefauthors}%
\unskip\
\newblock
\APACrefYearMonthDay{2007}{6}{}.
\newblock
{\BBOQ}\APACrefatitle {{Interprofessional relations and the emergence of a new
  profession: Software engineering in the United States, United Kingdom, and
  Canada}} {{Interprofessional relations and the emergence of a new profession:
  Software engineering in the United States, United Kingdom, and
  Canada}}.{\BBCQ}
\newblock
\APACjournalVolNumPages{Sociological Quarterly}{48}{3}{507--532}.
\PrintBackRefs{\CurrentBib}

\bibitem [\protect \citeauthoryear {%
{British Standards Institution}%
}{%
{British Standards Institution}%
}{%
{\protect \APACyear {2016}}%
}]{%
BritishStandardsInstitution2016BSSystems}
\APACinsertmetastar {%
BritishStandardsInstitution2016BSSystems}%
\begin{APACrefauthors}%
{British Standards Institution}.%
\end{APACrefauthors}%
\unskip\
\newblock
\APACrefYear{2016}.
\newblock
\APACrefbtitle {{BS 8611:2016 - Robots and robotic devices. Guide to the
  ethical design and application of robots and robotic systems}} {{BS 8611:2016
  - Robots and robotic devices. Guide to the ethical design and application of
  robots and robotic systems}}.
\PrintBackRefs{\CurrentBib}

\bibitem [\protect \citeauthoryear {%
Buiten%
}{%
Buiten%
}{%
{\protect \APACyear {2019}}%
}]{%
Buiten2019TowardsIntelligence}
\APACinsertmetastar {%
Buiten2019TowardsIntelligence}%
\begin{APACrefauthors}%
Buiten, M\BPBI C.%
\end{APACrefauthors}%
\unskip\
\newblock
\APACrefYearMonthDay{2019}{3}{}.
\newblock
{\BBOQ}\APACrefatitle {{Towards intelligent regulation of artificial
  intelligence}} {{Towards intelligent regulation of artificial
  intelligence}}.{\BBCQ}
\newblock
\BIn{} \APACrefbtitle {European Journal of Risk Regulation} {European journal
  of risk regulation}\ (\BVOL~10, \BPGS\ 41--59).
\newblock
\APACaddressPublisher{}{Cambridge University Press}.
\PrintBackRefs{\CurrentBib}

\bibitem [\protect \citeauthoryear {%
Burns%
\ \BBA {} Burns%
}{%
Burns%
\ \BBA {} Burns%
}{%
{\protect \APACyear {2019}}%
}]{%
Burns2019BeyondProfessions}
\APACinsertmetastar {%
Burns2019BeyondProfessions}%
\begin{APACrefauthors}%
Burns, E\BPBI A.%
\BCBT {}\ \BBA {} Burns, E\BPBI A.%
\end{APACrefauthors}%
\unskip\
\newblock
\APACrefYearMonthDay{2019}{}{}.
\newblock
{\BBOQ}\APACrefatitle {{Beyond Defining Professions}} {{Beyond Defining
  Professions}}.{\BBCQ}
\newblock
\BIn{} \APACrefbtitle {Theorising Professions} {Theorising professions}\
  (\BPGS\ 39--71).
\newblock
\APACaddressPublisher{}{Springer International Publishing}.
\newblock
\begin{APACrefDOI} \doi{10.1007/978-3-030-27935-6{\_}2} \end{APACrefDOI}
\PrintBackRefs{\CurrentBib}

\bibitem [\protect \citeauthoryear {%
Chatila%
\ \BBA {} Havens%
}{%
Chatila%
\ \BBA {} Havens%
}{%
{\protect \APACyear {2019}}%
}]{%
Chatila2019TheSystems}
\APACinsertmetastar {%
Chatila2019TheSystems}%
\begin{APACrefauthors}%
Chatila, R.%
\BCBT {}\ \BBA {} Havens, J\BPBI C.%
\end{APACrefauthors}%
\unskip\
\newblock
\APACrefYearMonthDay{2019}{}{}.
\newblock
{\BBOQ}\APACrefatitle {{The IEEE Global Initiative on Ethics of Autonomous and
  Intelligent Systems}} {{The IEEE Global Initiative on Ethics of Autonomous
  and Intelligent Systems}}.{\BBCQ}
\newblock
\BIn{} (\BPGS\ 11--16).
\PrintBackRefs{\CurrentBib}

\bibitem [\protect \citeauthoryear {%
Garvin%
}{%
Garvin%
}{%
{\protect \APACyear {1983}}%
}]{%
Garvin1983CanWork}
\APACinsertmetastar {%
Garvin1983CanWork}%
\begin{APACrefauthors}%
Garvin, D\BPBI A.%
\end{APACrefauthors}%
\unskip\
\newblock
\APACrefYearMonthDay{1983}{7}{}.
\newblock
{\BBOQ}\APACrefatitle {{Can Industry Self-Regulation Work?}} {{Can Industry
  Self-Regulation Work?}}{\BBCQ}
\newblock
\APACjournalVolNumPages{California Management Review}{25}{4}{37--52}.
\PrintBackRefs{\CurrentBib}

\bibitem [\protect \citeauthoryear {%
Godfrey%
}{%
Godfrey%
}{%
{\protect \APACyear {2009}}%
}]{%
Godfrey2009SystemsProfession}
\APACinsertmetastar {%
Godfrey2009SystemsProfession}%
\begin{APACrefauthors}%
Godfrey, P.%
\end{APACrefauthors}%
\unskip\
\newblock
\APACrefYearMonthDay{2009}{}{}.
\newblock
\APACrefbtitle {{'Systems engineering meets professional regulation:
  Establishing a recognised profession}} {{'Systems engineering meets
  professional regulation: Establishing a recognised profession}}\
  \APACbVolEdTR{}{\BTR{}}.
\PrintBackRefs{\CurrentBib}

\bibitem [\protect \citeauthoryear {%
Gorman%
\ \BBA {} Sandefur%
}{%
Gorman%
\ \BBA {} Sandefur%
}{%
{\protect \APACyear {2011}}%
}]{%
Gorman2011GoldenWork}
\APACinsertmetastar {%
Gorman2011GoldenWork}%
\begin{APACrefauthors}%
Gorman, E\BPBI H.%
\BCBT {}\ \BBA {} Sandefur, R\BPBI L.%
\end{APACrefauthors}%
\unskip\
\newblock
\APACrefYearMonthDay{2011}{8}{}.
\newblock
{\BBOQ}\APACrefatitle {{"Golden age," Quiescence, and Revival: How the
  sociology of professions became the study of knowledge-based work}} {{"Golden
  age," Quiescence, and Revival: How the sociology of professions became the
  study of knowledge-based work}}.{\BBCQ}
\newblock
\APACjournalVolNumPages{Work and Occupations}{38}{3}{275--302}.
\PrintBackRefs{\CurrentBib}

\bibitem [\protect \citeauthoryear {%
Holmes%
}{%
Holmes%
}{%
{\protect \APACyear {2000}}%
}]{%
Holmes2000FashioningProfession}
\APACinsertmetastar {%
Holmes2000FashioningProfession}%
\begin{APACrefauthors}%
Holmes, N.%
\end{APACrefauthors}%
\unskip\
\newblock
\APACrefYearMonthDay{2000}{7}{}.
\newblock
{\BBOQ}\APACrefatitle {{Fashioning a foundation for the computing profession}}
  {{Fashioning a foundation for the computing profession}}.{\BBCQ}
\newblock
\APACjournalVolNumPages{Computer}{33}{7}{97--98}.
\PrintBackRefs{\CurrentBib}

\bibitem [\protect \citeauthoryear {%
Holmes%
}{%
Holmes%
}{%
{\protect \APACyear {2010}}%
}]{%
Holmes2010TheProfession}
\APACinsertmetastar {%
Holmes2010TheProfession}%
\begin{APACrefauthors}%
Holmes, N.%
\end{APACrefauthors}%
\unskip\
\newblock
\APACrefYearMonthDay{2010}{7}{}.
\newblock
{\BBOQ}\APACrefatitle {{The future of the computing profession}} {{The future
  of the computing profession}}.{\BBCQ}
\newblock
\APACjournalVolNumPages{Computer}{43}{7}{}.
\PrintBackRefs{\CurrentBib}

\bibitem [\protect \citeauthoryear {%
Huising%
\ \BBA {} Silbey%
}{%
Huising%
\ \BBA {} Silbey%
}{%
{\protect \APACyear {2011}}%
}]{%
Huising2011GoverningRegulation}
\APACinsertmetastar {%
Huising2011GoverningRegulation}%
\begin{APACrefauthors}%
Huising, R.%
\BCBT {}\ \BBA {} Silbey, S\BPBI S.%
\end{APACrefauthors}%
\unskip\
\newblock
\APACrefYearMonthDay{2011}{3}{}.
\newblock
{\BBOQ}\APACrefatitle {{Governing the gap: Forging safe science through
  relational regulation}} {{Governing the gap: Forging safe science through
  relational regulation}}.{\BBCQ}
\newblock
\APACjournalVolNumPages{Regulation {\&} Governance}{5}{1}{14--42}.
\PrintBackRefs{\CurrentBib}

\bibitem [\protect \citeauthoryear {%
Lester%
}{%
Lester%
}{%
{\protect \APACyear {2016}}%
}]{%
Lester2016TheCommunities}
\APACinsertmetastar {%
Lester2016TheCommunities}%
\begin{APACrefauthors}%
Lester, S.%
\end{APACrefauthors}%
\unskip\
\newblock
\APACrefYearMonthDay{2016}{}{}.
\newblock
{\BBOQ}\APACrefatitle {{The development of self-regulation in four UK
  professional communities}} {{The development of self-regulation in four UK
  professional communities}}.{\BBCQ}
\newblock
\APACjournalVolNumPages{Professions and Professionalism}{6}{1}{1--14}.
\PrintBackRefs{\CurrentBib}

\bibitem [\protect \citeauthoryear {%
Mitrou%
}{%
Mitrou%
}{%
{\protect \APACyear {2019}}%
}]{%
Mitrou2019DataIntelligence-Proof}
\APACinsertmetastar {%
Mitrou2019DataIntelligence-Proof}%
\begin{APACrefauthors}%
Mitrou, L.%
\end{APACrefauthors}%
\unskip\
\newblock
\APACrefYearMonthDay{2019}{6}{}.
\newblock
{\BBOQ}\APACrefatitle {{Data Protection, Artificial Intelligence and Cognitive
  Services: Is the General Data Protection Regulation (GDPR) ‘Artificial
  Intelligence-Proof’?}} {{Data Protection, Artificial Intelligence and
  Cognitive Services: Is the General Data Protection Regulation (GDPR)
  ‘Artificial Intelligence-Proof’?}}{\BBCQ}
\newblock
\APACjournalVolNumPages{SSRN Electronic Journal}{}{}{}.
\PrintBackRefs{\CurrentBib}

\bibitem [\protect \citeauthoryear {%
Petit%
}{%
Petit%
}{%
{\protect \APACyear {2017}}%
}]{%
Petit2017LawImplications}
\APACinsertmetastar {%
Petit2017LawImplications}%
\begin{APACrefauthors}%
Petit, N.%
\end{APACrefauthors}%
\unskip\
\newblock
\APACrefYearMonthDay{2017}{3}{}.
\newblock
{\BBOQ}\APACrefatitle {{Law and Regulation of Artificial Intelligence and
  Robots - Conceptual Framework and Normative Implications}} {{Law and
  Regulation of Artificial Intelligence and Robots - Conceptual Framework and
  Normative Implications}}.{\BBCQ}
\newblock
\APACjournalVolNumPages{SSRN Electronic Journal}{}{}{}.
\PrintBackRefs{\CurrentBib}

\bibitem [\protect \citeauthoryear {%
\APACcitebtitle {{Principles of Robotics - EPSRC website}}}{%
\APACcitebtitle {{Principles of Robotics - EPSRC website}}}{%
{\protect \APACyear {2010}}%
}]{%
2010PrinciplesWebsite}
\APACinsertmetastar {%
2010PrinciplesWebsite}%
\APACrefbtitle {{Principles of Robotics - EPSRC website}.} {{Principles of
  Robotics - EPSRC website}.}
\newblock
\APACrefYearMonthDay{2010}{}{}.
\newblock
\begin{APACrefURL}
  \url{https://epsrc.ukri.org/research/ourportfolio/themes/engineering/activities/principlesofrobotics/}
  \end{APACrefURL}
\PrintBackRefs{\CurrentBib}

\bibitem [\protect \citeauthoryear {%
Vinarova%
\ \BBA {} Mihova%
}{%
Vinarova%
\ \BBA {} Mihova%
}{%
{\protect \APACyear {2010}}%
}]{%
Vinarova2010TheInformatician}
\APACinsertmetastar {%
Vinarova2010TheInformatician}%
\begin{APACrefauthors}%
Vinarova, J.%
\BCBT {}\ \BBA {} Mihova, P.%
\end{APACrefauthors}%
\unskip\
\newblock
\APACrefYearMonthDay{2010}{9}{}.
\newblock
{\BBOQ}\APACrefatitle {{The profession “medical informatician”}} {{The
  profession “medical informatician”}}.{\BBCQ}
\newblock
\APACjournalVolNumPages{Medical and Health Science Journal}{3}{3}{67--71}.
\PrintBackRefs{\CurrentBib}

\bibitem [\protect \citeauthoryear {%
Winfield%
}{%
Winfield%
}{%
{\protect \APACyear {2019}}%
}]{%
Winfield2019EthicalAI}
\APACinsertmetastar {%
Winfield2019EthicalAI}%
\begin{APACrefauthors}%
Winfield, A.%
\end{APACrefauthors}%
\unskip\
\newblock
\APACrefYearMonthDay{2019}{2}{}.
\newblock
\APACrefbtitle {{Ethical standards in robotics and AI}} {{Ethical standards in
  robotics and AI}}\ (\BVOL~2)\ (\BNUM~2).
\newblock
\APACaddressPublisher{}{Nature Publishing Group}.
\PrintBackRefs{\CurrentBib}

\end{thebibliography}

\end{document}